\documentclass[preprintnumbers,prd,twocolumn,floatfix,nofootinbib,
superscriptaddress]{revtex4}
\usepackage{ads}
\usepackage[colorlinks,linkcolor=blue,citecolor=blue]{hyperref}
\pdfoutput=1
\usepackage[utf8]{inputenc}

\usepackage{graphicx}
\usepackage{dcolumn}
\usepackage{natbib}

\begin{document}

\title{ N-body simulations, halo mass functions, and halo density profile   in 
 $f(T)$ gravity}

\author{Yiqi Huang}
\email{yq\_huang@sjtu.edu.cn}
\affiliation{Department of Astronomy, Shanghai Jiao Tong University, Shanghai 
200240, China}

\author{Jiajun Zhang}
\email{jjzhang@shao.ac.cn}
\affiliation{Shanghai Astronomical Observatory, Chinese Academy of Sciences, 
Shanghai 200030, China}

\author{Xin Ren}
\affiliation{Department of Astronomy, School of Physical Sciences,
University of Science and Technology of China, Hefei, Anhui 230026, China}
\affiliation{CAS Key Laboratory for Researches in Galaxies and Cosmology, School 
of Astronomy and Space Science, University of Science and Technology of China, 
Hefei, Anhui 230026, China}

\author{Emmanuel N. Saridakis}
\affiliation{National Observatory of Athens, Lofos Nymfon, 11852 Athens, Greece}
\affiliation{Department of Astronomy, School of Physical Sciences,
University of Science and Technology of China, Hefei, Anhui 230026, China}
\affiliation{CAS Key Laboratory for Researches in Galaxies and Cosmology, School 
of Astronomy and Space Science, University of Science and Technology of China, 
Hefei, Anhui 230026, China}

\author{Yi-Fu Cai}
\affiliation{Department of Astronomy, School of Physical Sciences,
University of Science and Technology of China, Hefei, Anhui 230026, China}
\affiliation{CAS Key Laboratory for Researches in Galaxies and Cosmology, School 
of Astronomy and Space Science, University of Science and Technology of China, 
Hefei, Anhui 230026, China}


\begin{abstract}
 
We perform N-body simulations for $f(T)$ gravity using the ME-Gadget code, in 
order to investigate for the first time the structure formation 
process  in detail. Focusing on the power-law model, and considering the 
model-parameter   to be consistent within 1$\sigma$ with all other cosmological 
datasets (such as SNIa, BAO, CMB, CC), we show  that  there are clear 
observational differences between $\Lambda$CDM cosmology and $f(T)$ gravity, 
due to the 
modifications brought about the latter in the Hubble function evolution and the 
effective Newton's constant. We extract the matter density 
distribution, matter power spectrum, counts-in-cells, halo mass function and 
excess surface density (ESD) around low density positions (LDPs) at present time. 
Concerning the  matter power spectrum we find a difference from $\Lambda$CDM 
scenario, which is attributed to about $2/3$ to the  different expansion and to 
about $1/3$ to the effective gravitational constant. Additionally, we find a 
difference in the cells, which is significantly  
larger than the Poisson error, which may be distinguishable with weak-lensing 
reconstructed mass maps. Moreover, we show that there are different
 massive halos with mass $M>10^{14}M_{\odot}/h$, which may be 
distinguishable with statistical measurements of cluster number counting, and 
we  find that the ESD around LDPs is mildly 
different. In conclusion, high-lighting possible smoking guns, we show that 
large scale structure can indeed lead us to distinguish General Relativity and 
$\Lambda$CDM cosmology from $f(T)$ gravity.

\end{abstract}
\maketitle

\section{Introduction}
Since the discovery of   late-time cosmic acceleration at the end of 
last century 
 \citep{SupernovaSearchTeam:1998fmf, 
SupernovaCosmologyProject:1998vns,SupernovaSearchTeam:2003cyd}, a number 
of 
observations of various kinds
 \citep{SDSS:2003eyi, Barris:2003dq,  SupernovaCosmologyProject:2003dcn, 
SupernovaSearchTeam:2004lze} have confirmed 
its existence.
The simplest paradigm to 
explain most of these observations is the 
$\Lambda$CDM  one, in which the source of 
acceleration is the  
cosmological constant $\Lambda$ within the context of general relativity (GR) 
and with the additional presence of the Cold Dark Matter (CDM)
sector  \citep{2000IJMPD...9..373S}. However, the possibility of a dynamical 
nature, as well as the need to describe also the initial (inflationary) 
accelerating phase \cite{Olive:1989nu}, lead to two ways of modification of the 
above concordance model.  The first direction is to  introduce the concept of 
dark 
energy within general relativity \cite{Copeland:2006wr,Cai:2009zp}, while the 
second way is 
to construct modifications of gravity which can offer the new degrees of 
freedom required to describe cosmic acceleration 
\cite{CANTATA:2021ktz,Capozziello:2011et}. Note that the second 
direction has the advantage of being closer to a quantum description of gravity
\cite{Addazi:2021xuf} as well as alleviating the possible tension 
\cite{Abdalla:2022yfr}  between local Hubble constant observations 
 \citep{2019ApJ...876...85R} and Cosmic Microwave Background 
(CMB) 
estimations  \citep{Planck:2018vyg}.
 
There are many ways that one can follow in order to construct modified and 
extended theories of gravity. The most used one is to start from the 
Einstein-Hilbert Lagrangian, namely from the curvature formulation of gravity, 
and extend it in various ways, resulting to  $f(R)$ gravity 
\cite{Starobinsky:1980te,Capozziello:2002rd,DeFelice:2010aj,Nojiri:2010wj}, 
$f(G)$ gravity \cite{Nojiri:2005jg, DeFelice:2008wz}, Lovelock 
gravity \cite{Lovelock:1971yv}, Weyl theory
\cite{Mannheim:1988dj}   Galileon gravity 
\cite{Nicolis:2008in, Deffayet:2009wt}, etc. 
An interesting alternative is to start from the equivalent, torsional 
formulation of gravity \cite{Cai:2015emx}, and extend it in many ways, such as 
in   $f(T)$ gravity \cite{Bengochea:2008gz, Linder:2010py, Chen:2010va}, in  
$f(T,T_G)$ gravity \cite{Kofinas:2014owa}, in symmetric 
teleparallel gravity
\cite{BeltranJimenez:2019tme}, etc. The cosmological 
applications of $f(T)$      gravity have been proven 
to be very interesting  
 \citep{Cai:2015emx, Ren:2021uqb, Bahamonde:2021gfp} and the theory 
has also been confronted to solar-system observations 
\cite{Iorio:2012cm,Iorio:2015rla,Farrugia:2016xcw}, gravitational waves data 
\cite{Cai:2018rzd,Nunes:2018evm,Nunes:2019bjq}, as well as cosmological data 
from  Supernovae type Ia
data  (SNIa),  Cosmic Microwave Background (CMB),
Baryonic Acoustic Oscillations (BAO), and $f\sigma_8$ observations
\cite{Wu:2010mn,Cardone:2012xq,Nesseris:2013jea, 
Basilakos:2016xob, Nunes:2016plz,Nunes:2018xbm,Basilakos:2018arq,
Xu:2018npu,  Cai:2019bdh,Anagnostopoulos:2019miu,Ren:2021tfi,Ren:2022aeo}.

Since General Relativity (GR)  has passed very precise tests in the Solar 
System  \citep{turyshev2008experimental}, any modified gravity   should be 
very close to GR in these scales. Therefore, the 
screening mechanism is introduced to provide the link between deviations from 
GR at cosmological scales and local consistency with GR. In 
particular, screening mechanisms  \citep{2018LRR....21....1B} are   
prevalent extensions of GR over the last decade, and they utilize non-linear 
dynamics to introduce new scalar fields that couple to gravity (these scalar 
fields can behave very differently between solar-system and cosmological 
scales).  

In order to confront modified gravity theories with various kinds of 
observational data, we need to extract their precise predictions,  
from cosmic expansion to structure formation. N-body simulation has been proved 
to be a powerful tool to study large scale structure, and its 
application to various modified theories of gravity   has undergone rapid 
developments through the past decade  \citep{2020MNRAS.497.1885H}. 
Additionally, 
with the application of large galaxy surveys, weak 
lensing  \citep{2001PhR...340..291B} has been proven a strong method to probe 
the matter distribution of the large scale structure, while it also provides a 
way to test modified gravity theories. Within this field, cosmic 
voids  \citep{2014PhRvD..90j3521C} have obtained much 
attention  \citep{zhang2020cubic} due to the fact that they are less affected 
by 
baryonic physics and nonlinear evolution compared to dark matter halos.

In 
this work we are interested in performing   N-body simulations in the case of 
$f(T)$ gravity. For this purpose we use the ME-Gadget 
code  \citep{megadget,zhang2019ApJ}. Furthermore, we will analyze the 
ESD around low density positions in $f(T)$ gravity and we will extract the halo 
mass functions. The paper is organized as follows:
In Section \ref{sec:model} we briefly provide  the cosmological equations of 
  $f(T)$ gravity at the background and perturbation levels. In Section 
 \ref{sec:nbody} we introduce the 
ME-Gadget code  \citep{megadget} and we describe the simulation steps we 
utilize. The results of our simulations  are presented and discussed in 
Section \ref{sec:result}. Finally, Section 
\ref{sec:summary} is devoted to summary and conclusions.

\section{ $f(T)$ gravity and cosmology}
\label{sec:model}

Let us briefly present $f(T)$ gravity and apply it in a cosmological framework. 
We start from    teleparallel gravity, in which one uses torsion instead 
of curvature to describe gravity, and where for convenience one uses the 
tetrad fields $e^{A}{}_{\mu}$, namely the four orthonormal vectors in the 
tangent space, as the dynamical varables (Greek  and Latin indices are  
respectively used for the coordinate and
tangent space). Note that the metric is related to the tetrad field through
$g_{\mu \nu}:g_{\mu \nu}=\eta_{A B} e^{A}{}_{\mu} e^{B}{}_{\nu}$,
with $\eta_{AB}={\rm diag} 
(1,-1,-1,-1)$.
One introduces the
Weitzenb$\ddot{\text{o}}$ck connection as 
$\hat{\Gamma}^{\lambda}{}_{\mu \nu} \equiv e_{A}{}^{\lambda} \partial_{\nu} 
e^{A}{}_{\mu}$, and thus the corresponding torsion tensor is
\begin{equation}
T^{\lambda}{}_{\mu \nu} \equiv \hat{\Gamma}^{\lambda}{}_{\nu 
\mu}-\hat{\Gamma}^{\lambda}{}_{\mu \nu}=e_{A}^{}{\lambda}\left(\partial_{\mu} 
e^{A}{}_{\nu}-\partial_{\nu} e^{A}{}_{ \mu}\right) ,    
\end{equation}
while  the torsion   scalar $T$ reads as
\begin{equation}
T \equiv \frac{1}{4} T^{\rho \mu \nu} T_{\rho \mu \nu}+\frac{1}{2} T^{\rho \mu 
\nu} T_{\nu \mu \rho}-T_{\rho \mu}{}{}^{\rho} T^{\nu \mu}{}{}_{\nu}.
\end{equation}
Since the Ricci scalar $R$ corresponding to the torsionless 
Levi-Civita connection, and $T$, differ by only a boundary term, their 
corresponding use as Lagrangians leads to equivalent theories, namely to GR and 
the teleparallel equivalent of general relativity (TEGR), respectively.

One can now construct $f(T)$ gravity by extending the action of TEGR, writing 
the action
\begin{equation}
 S=\int d^4x\frac{e}{16\pi G}[T+f(T)+L_m+L_r],   
\end{equation}
where $e=det\left(e^{A}{}_{\mu}\right)=\sqrt{-g}$,  and $L_m$ and $L_r$ 
represent the Lagrangians of matter and radiation sector respectively. In order 
to proceed to the cosmological application, we impose the flat 
Friedmann-Robtson-Walker (FRW) metric 
\begin{equation}
d s^{2}=d t^{2}-a^{2}(t) \delta_{ij} d x^{i} d x^{j},
\end{equation}
where $a(t)$ is the scale factor. The corresponding tetrad is  
$e_{\mu}^A={\rm
diag}(1,a,a,a)$, and one can find that $T = -6H^2$. Hence, 
the modified Friedmann equations in $f(T)$ cosmology can be extracted as
\begin{equation}\label{eq:friedmann1}
H^{2} = \frac{8 \pi G}{3} (\rho_{m}+\rho_{r})-\frac{f(T)}{6}+\frac{T f_{T}}{3} 
\end{equation}
\begin{equation}\label{eq:friedmann2}
\dot{H} = -\frac{4 \pi G\left(\rho_{m}+P_{m}+\rho_{r}+P_{r}\right)}{1+f_{T}+2 T 
f_{T T}},
\end{equation}
with   $f_{T}\equiv\partial f/\partial T$, 
$f_{TT}\equiv\partial^{2} 
f/\partial T^{2}$, and  
where
$\rho_m,\rho_r$ and $P_m,P_r$ represent the energy densities and pressures for 
matter and radiation respectively. Therefore, the extra   
terms of gravitational origin  can be considered 
as an effective    dark-energy component.

We can express the background cosmological equations in a more convenient way, 
by introducing the 
function $E(z)$ as
\begin{equation}
E^2(z)\equiv \frac{H^2(z)}{H^2_0}=\frac{T(z)}{T_0},
\end{equation}
where the subscript ``0'' denotes the value of a quantity at present time. 
Then, the first Friedmann equation 
becomes
\begin{equation}
E^2(z,r)=\Omega_{m0}(1+z)^3+\Omega_{r0}(1+z)^4+\Omega_{F0}y(z,r),
\end{equation}
with 
\begin{equation}
y(z,r)=\frac{1}{T_0\Omega_{F0}}[f-2Tf_T],
\end{equation}
and where $\Omega_{F0}\equiv1-\Omega_{m0}-\Omega_{r0}$. 

Let us now examine the linear matter perturbation level. In particular, 
  the   linear growth rate under the   $f(T)$ 
gravity, is determined by the equation  
\cite{Nesseris:2013jea,Anagnostopoulos:2019miu,Ren:2021tfi}
\begin{equation}
\delta^{\prime\prime}+\delta^{\prime}\left(\frac{H^\prime}{H}+\frac{3}{a}
\right)-\frac { G_ { eff}}{G } 
\frac{\delta}{a^2}\left(\frac{3\Omega_mH_0^2}{2H^2a^3}\right)=0,
\end{equation}
where $\delta\equiv\delta\rho_m/\rho_m$ is the matter overdensity and
with $\delta^{\prime}\equiv d\delta/da$ and $\delta^{\prime\prime}\equiv 
d^2\delta/da^2$. We mention that in the above expression the effect of $f(T)$ 
gravity is quantified   into the effective Newtonian constant, which according 
to the Poisson equations is found to be 
\cite{Zheng:2010am,Nesseris:2013jea,Anagnostopoulos:2019miu,Yan:2019gbw}: 
\begin{equation}\label{eq:geff}
G_{eff}=\frac{G }{1+f_T}.
\end{equation}

In this work we focus on the widely-used power-law $f(T)$ model
 \citep{2009PhRvD..79l4019B}, namely:
\begin{equation}
f(T)=T+\alpha(-T)^b,
\end{equation}
with $b$ the free model parameter, and where the parameter $\alpha$ is 
related to     $\Omega_{F0}$  through 
\begin{equation}
\alpha=(6H^2_0)^{1-b}\frac{\Omega_{F0}}{2b-1}.
\end{equation}
Thus, the effective Newton's constant in this specific model becomes:
\begin{equation}
G_{eff}(z)=\frac{G_N}{1+\frac{b\Omega_{F0}}{(1-2b)E^{2(1-b)}}}. 
\end{equation}
Finally, we mention  that this model recovers $\Lambda$CDM
 scenario for $b=0$, i.e. 
$T+f(T)=T-2\Lambda$, with $\Lambda=3\Omega_{F0}H_0^2$ and 
$\Omega_{F0}=\Omega_{\Lambda0}$.
Moreover, cosmological data form various
origins lead to the constraint     b$\in$[-0.29,0.26] and 
$\Omega_{m0}\in$[0.24,0.308]
\cite{Wu:2010mn,Cardone:2012xq,Nesseris:2013jea, 
Basilakos:2016xob, Nunes:2016plz,Nunes:2018xbm,Basilakos:2018arq,
Xu:2018npu,  Cai:2019bdh,Anagnostopoulos:2019miu}.

\section{N-body simulations}\label{sec:nbody}

In this section we present the method of N-body simulations, in order to apply 
it to the case of the power-law $f(T)$ gravity. 
Screening mechanism is a widely adopted mechanism for modified gravity 
theories \citep{2010fRreview} in order to be consistent with local GR tests. 
There are several different N-body simulation codes implementing this 
mechanism for modified gravity 
  \citep{li2012JCAP,barreira2015JCAP,bose2017JCAP}. In the following 
we will use    the 
ME-Gadget code  \citep{megadget,zhang2019ApJ} \footnote{ 
https://github.com/liambx/ME-Gadget-public} in order to test the effect of 
screening phenomenologically, using a
periodic box of size 512 Mpc/h and $512^3$ particles. 

The ME-Gadget code is a 
modified version of the public available N-body simulation code 
Gadget2 \citep{gadget2}. The original modification algorithm was proposed in 
\citet{baldi2010MNRAS} for coupled dark energy models. ME-Gadget was designed 
to 
use tabulated text files to include the change of $H(z)$ and $G_{eff}(z,k)$, 
which makes it perfect to study the structure formation of $f(T)$ gravity 
model. 
The pre-initial condition is prepared with CCVT algorithm \citep{ccvt} and the 
initial condition is calculated with second-order Lagrangian Perturbation 
Theory 
using a modified version of 2LPTic 
code \citep{2lpt}\footnote{https://github.com/liambx/Simp2LPTic}. Finally, the 
halos are 
identified using Amiga Halo Finder \citep{ahf}.

We impose a free parameter $k_{screen}$, and 
therefore for scales $k<k_{screen}$ we adopt the $G_{eff}(z)$ arising from 
$f(T)$ gravity, namely expression (\ref{eq:geff}), while  for scales 
$k>k_{screen}$ we adopt the usual Newtonian gravitational constant.
We choose three $k_{screen}$ values for 
the  simulations, namely 0.025, 0.05 and 0.1.
Concerning the $f(T)$ gravity model parameter $b$, without loss of generality 
we consider the extreme values according to cosmological constraints, i.e. 
$b=0.20579\equiv b_1$ and $b=-0.2371\equiv b_2$, and for completeness we 
consider also the value $b=0\equiv b_0$ which corresponds to $\Lambda$CDM 
cosmology. Finally we set $\Omega_{m0}=0.2917,h=0.7324$ 
 \citep{Planck:2018vyg}. In summary,  in our simulations
  we have two free 
parameters, $b$ and $k_{screen}$, and the corresponding models are presented in 
 Table \ref{tab:table1}.

\begin{table} 
\begin{tabular}{cccc}
\hline \hline
  &$k_{screen}=0.025\ $&$k_{screen}=0.05\ $&$k_{screen}=0.1\ $\\ \hline
 (b0): b=0&-&-&- \\
 (b1): b=0.20579    &b1k0025&b1k005&b1k01\\
 (b2): b=-0.2371 &b2k0025&b2k005&b2k01\\
 \hline\hline
\end{tabular}
\caption{\label{tab:table1}Model abbreviations  corresponding to different 
simulations.}
\end{table}

\begin{figure}[ht]
    \includegraphics[width=0.45\textwidth]{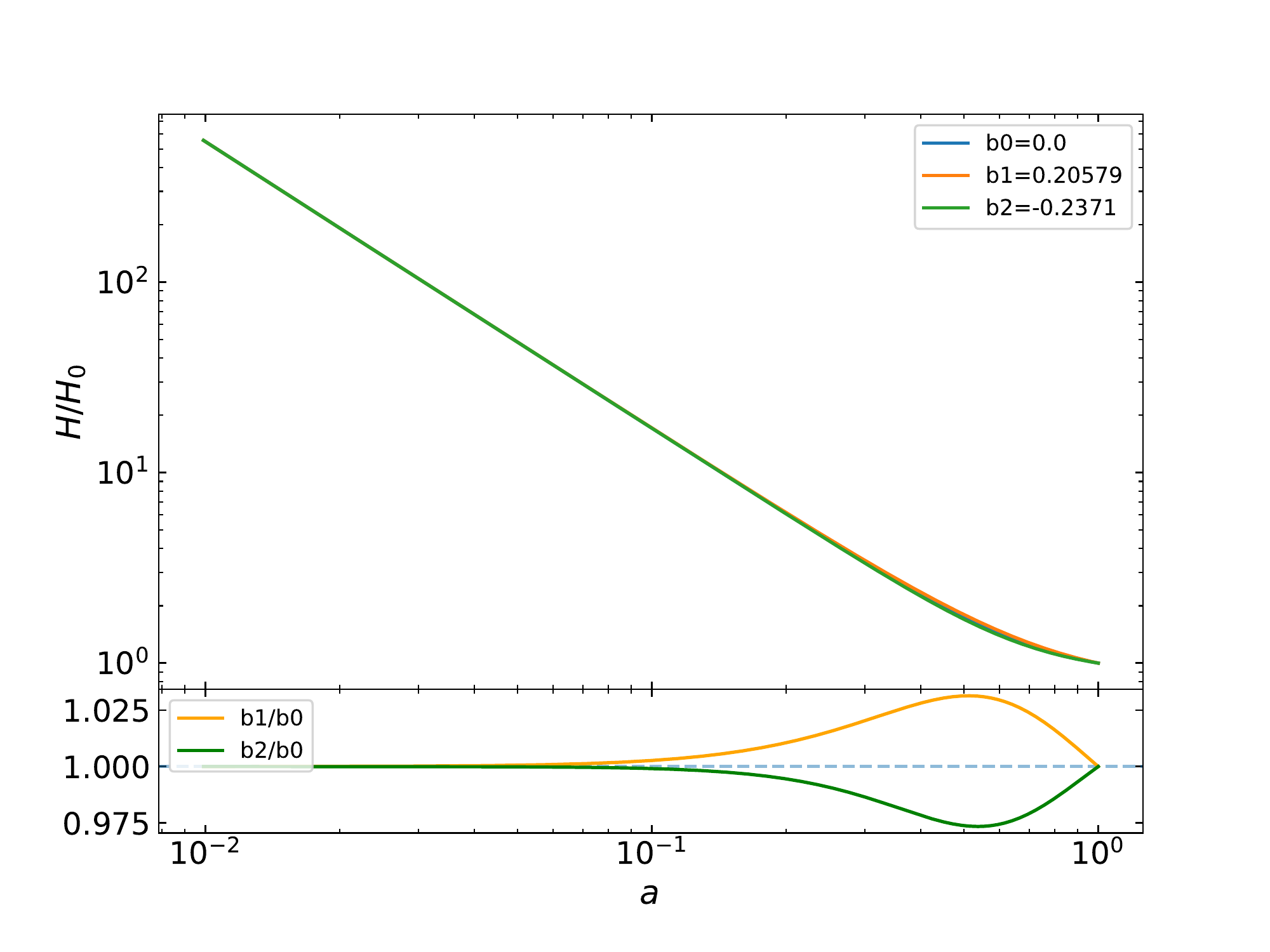}
    \caption{\label{Fig:Hz}{\it{ In the main graph we depict the Hubble 
function evolution in terms of the scale factor $a$ (normalized to $a=1$ at 
present), in the case of b0, b1 and b2 $f(T)$ gravity models, while in the 
bottom panel we present the ratio of the Huble function between b1, b2 models 
and b0, i.e. $\Lambda$CDM cosmology.}}}
\end{figure}
\begin{figure}[ht]
    \includegraphics[width=0.45\textwidth]{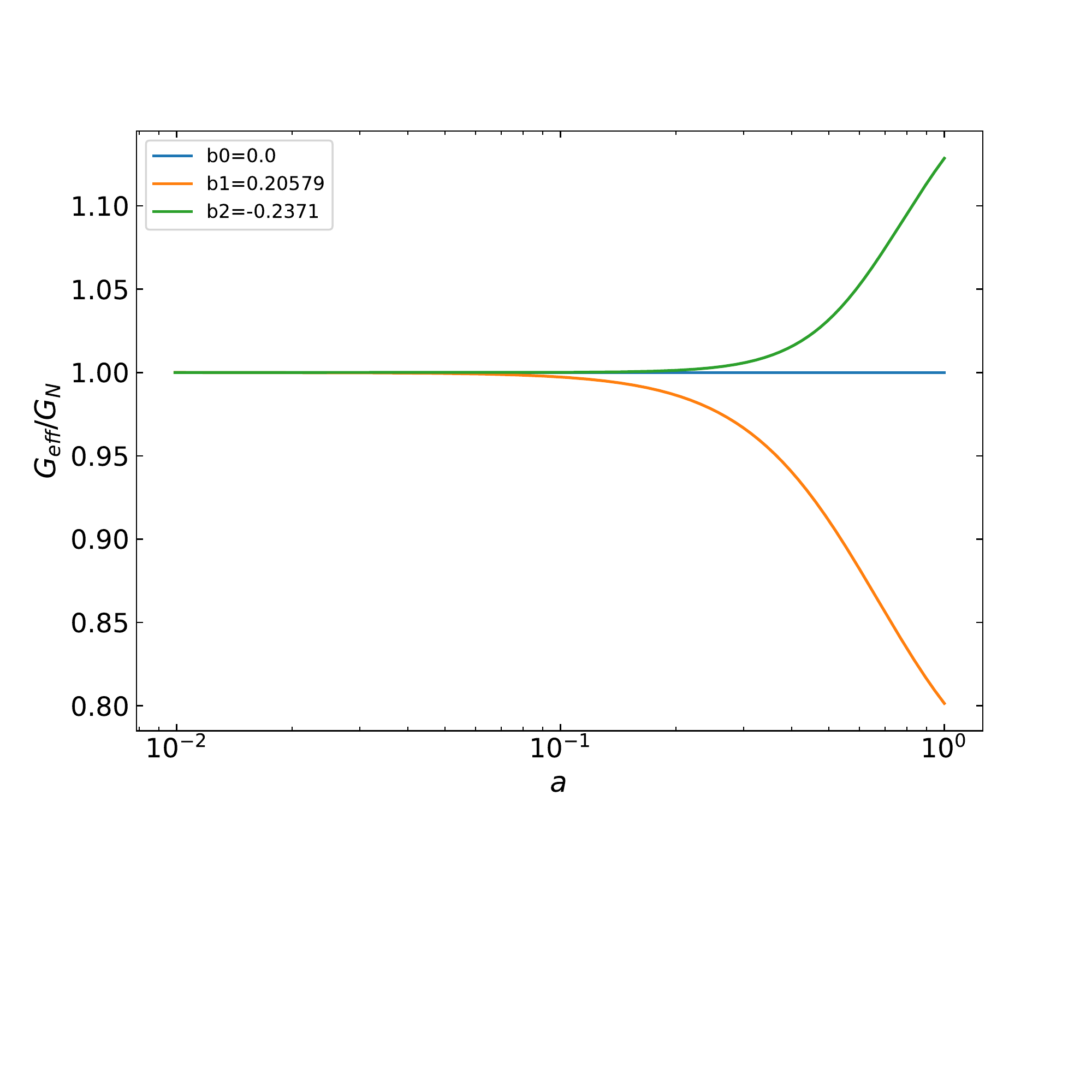}
    \vspace{-2cm}
    \caption{\label{Fig:Geff}
    {\it{ 
The evolution of the ratio  $G_{eff}/G$ in terms of the scale factor $a$ 
(normalized to $a=1$ at 
present), in the case of b0, i.e. $\Lambda$CDM cosmology, and for b1 and b2 
$f(T)$ gravity models.}}}
\end{figure}
\begin{figure*}
    \centering
    \includegraphics[width=0.95\textwidth]{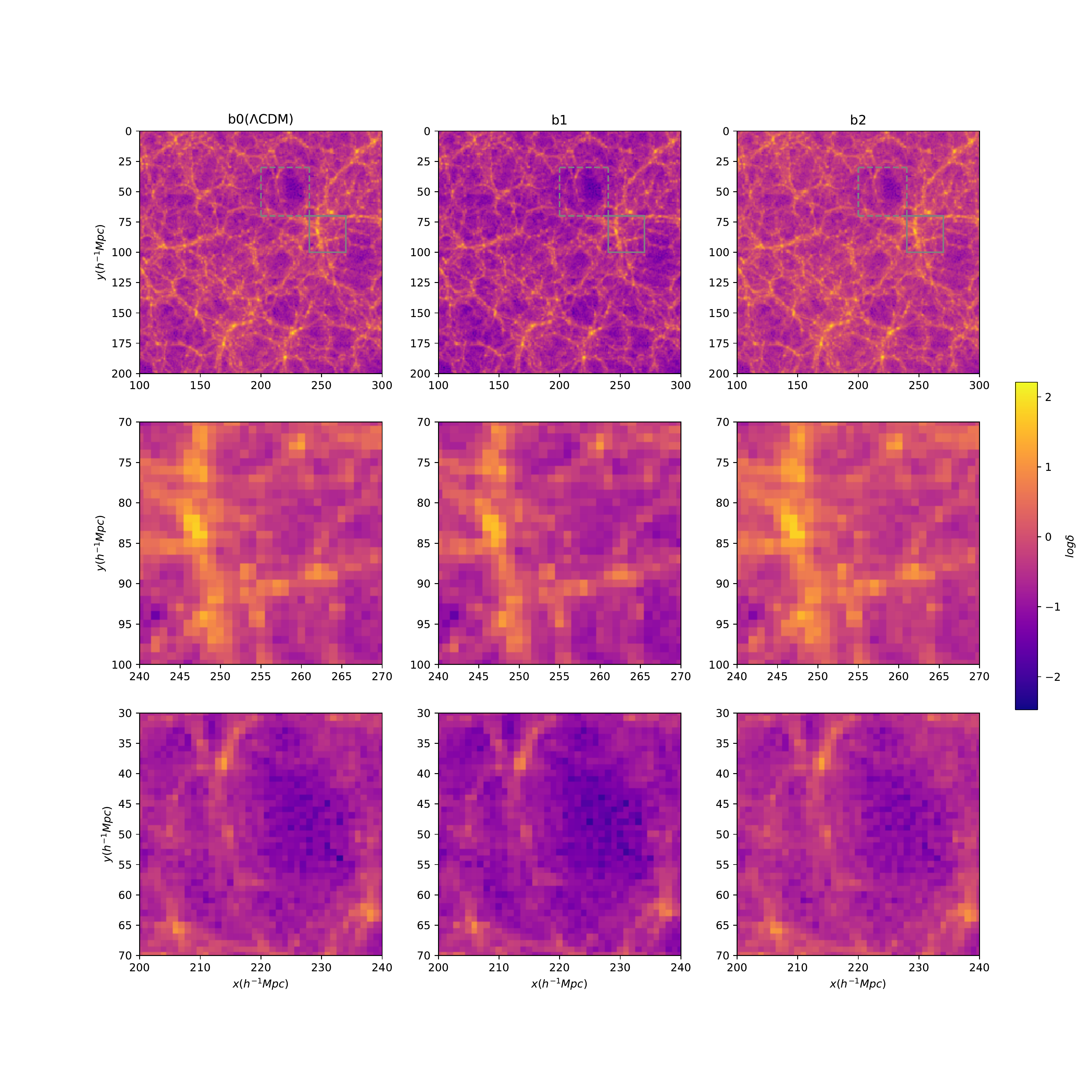}
    \vspace{-1cm}
    \caption{\label{matter}
    {\it{
    The dark matter density distributions for b0 model, i.e. $\Lambda$CDM 
cosmology, 
and for b1 and b2 
$f(T)$ gravity models, in 2D maps at present time $a=1$. The upper panel shows 
a  
$200Mpc/h$-wide simulation box by projecting $20Mpc/h$-slices along $z$-axis, 
while the lower panels are the zoom-in results, for an over-dense region 
(second 
row) and an under-dense region (third row), shown as the solid-line and 
dotted-line squares plotted in the first panel respectively.}}} 
\end{figure*}
 
In Fig.~\ref{Fig:Hz} we depict the  Hubble function evolution for the three 
examined cases of $f(T)$ gravity, and the corresponding deviation from 
  $\Lambda$CDM scenario. Note that the deviation is not more than 
$3\%$, and the largest difference is realized at scale factors around 
$a\sim0.5$ 
or around  redshift $z=1$. On the other hand, both at high redshift and 
redshift close to 0, the difference is negligible. 
Additionally, in Fig.~\ref{Fig:Geff} we present the evolution of the effective 
gravitational constant $G_{eff}$ as a function of the
scale factor. As we observe, $G_{eff}$ starts to deviate from $G$ 
 more than $5\%$ around $a=0.5$, too. Furthermore, the 
difference is increasing as $a$ approaches the present value $a=1$. Since 
structure formation process is cumulative over time, we deduce that the most 
significant difference of all observables is expected to be measured at 
$a=1$, i.e at redshift $z=0$.

In summary, we perform  N-body simulations and our purpose 
  is   to identify which observable 
may be possible to quantify the effect of $f(T)$ gravity.

\section{Results}
\label{sec:result}

In this section we present the obtained results of  N-body simulations in 
detail. In particular, we will examine the matter density distribution and the 
matter power spectrum, the  redshift-space distortion effect, the  halo mass 
function,
the halo density profile, and the low-density 
position (LDP) lensing.

\subsection{Matter density distribution}

For a straightforward comparison of the simulation, we use  the Pylians python 
library  \citep{2018ApJ...866..135V} 
\footnote{https://github.com/franciscovillaescusa/Pylians} to calculate the 
dark-matter over-density  $\delta\equiv\delta\rho_m/\rho_m$.  We choose  a 
$20Mpc/h$ thick slice along the $z$-axis at $a=1$ for illustration, and  in 
Fig.~\ref{matter} we   show the logrithmic dark-matter over-density 
distribution, focusing  on the results for $k_{screen}=0.05$. Since we use 
the same initial condition for each model, their structures are almost   the 
same. 

The first column depicts the $\Lambda$CDM scenario, the second column   the 
$b1$ model, and the third column shows the $b2$ model. In the first row, we 
present  the density distribution over the whole simulation box. Moreover, in 
the second row we     zoom-in  for one over-dense region, while in 
the third row we show the neighboring low-dense region. Comparing to the 
$\Lambda$CDM scenario, we see that there have been faster (slower) cosmic 
expansion and smaller (larger) gravitational constant in b1(b2) model. 
Therefore, it is expected that the structure growth in b1(b2) case should be 
less (more) strong than that in the $\Lambda$CDM paradigm. Qualitatively, the 
difference can be seen from the density distribution, which motivates us to 
observe the quantitative difference in more detail.

\subsection{Matter power spectrum}

\begin{figure}
    \centering
    \includegraphics[width=0.48\textwidth]{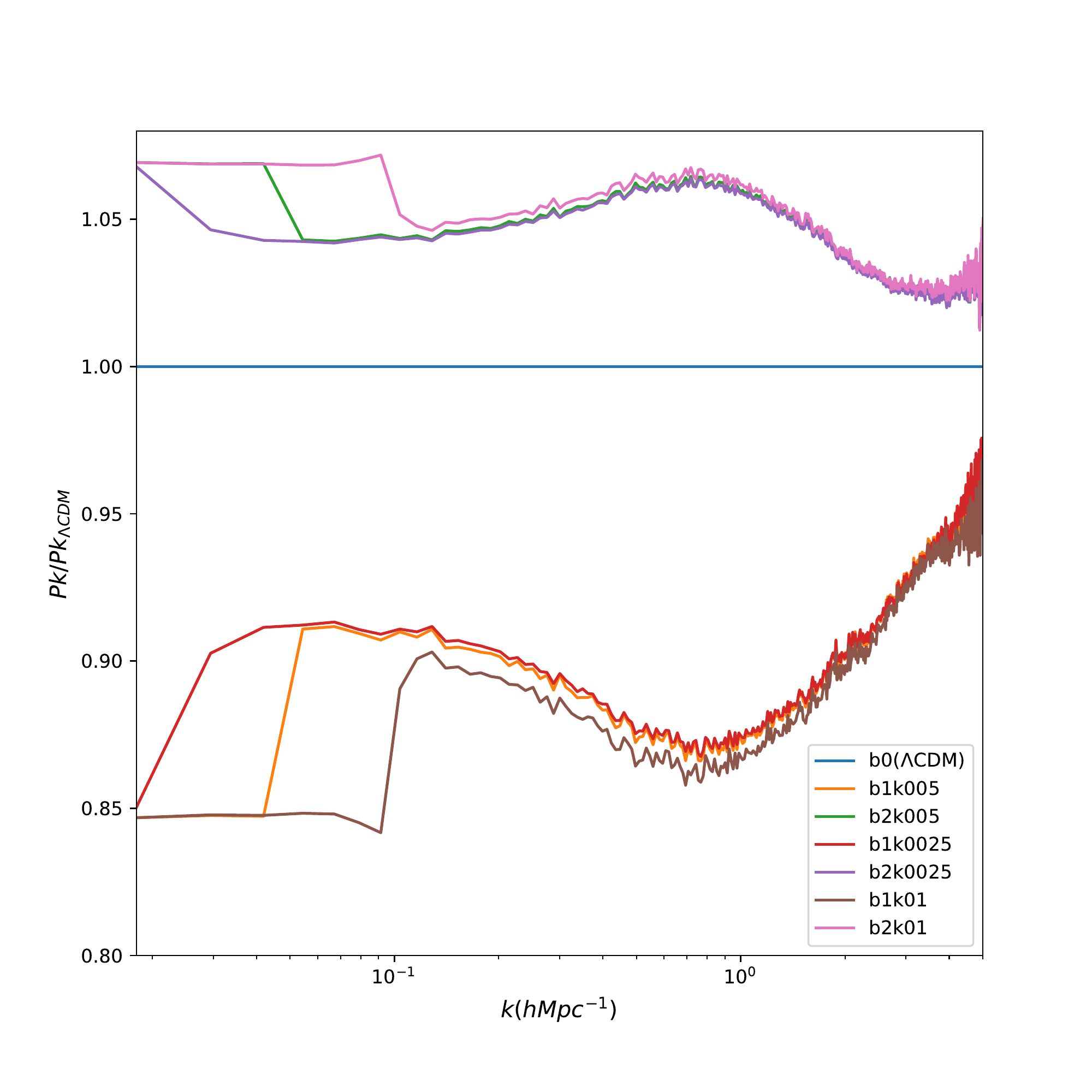}
    \caption{\label{Fig:pk}
    {\it{The dark-matter power spectrum ratio for b0 model, i.e. $\Lambda$CDM 
cosmology, and for b1 and b2 
$f(T)$ gravity models, for 3 different screening scales $k_{screen}=0.025, 
0.05, 0.1 \ hMpc^{-1}$ (see Table \ref{tab:table1}), at   present time 
$a=1$.}}}
\end{figure}

To quantitatively compare the  difference between b0, b1 and b2 models, we 
calculate their matter power spectrum at $a=1$, and we present the results  in 
Fig. \ref{Fig:pk}. The power spectrum is the correlation within a certain 
density field in $k$-space. We can clearly notice two phenomena in 
this figure. Firstly, the matter power spectrum in b1(b2) is 
smaller (larger) than that of $\Lambda$CDM scenario, and secondly there is a 
sharp jump of the matter power spectrum
ratio at $k=0.025,0.5,0.1(hMpc^{-1})$, 
which is exactly related to the screening scale we have set for the 
simulations. 
The difference of matter power spectrum at all scales is caused by the 
different expansion history and the sharp jump, provided the understanding 
of the effects of the effective gravitational constant. By 
looking at the ratio of matter power spectrum at scales smaller than the 
screening scale ($k<k_{screen}$), we can understand that quantitatively the 
change of $H(z)$ results to a $\sim5\%$ power-spectrum increase for b2 model 
and to a $\sim10\%$ power-spectrum decrease for b1 model. By looking at the 
ratio of matter power spectrum at scales larger than the screening scale 
($k>k_{screen}$), the change of $G_{eff}(z)$ results to an additional 
$\sim2.5\%$ power spectrum increase for b2 and $\sim5\%$ power spectrum 
decrease 
for b1 model. The percent-level difference is actually distinguishable, 
nevertheless the problem is that  the value of the power spectrum is also 
determined by $\sigma_8$, which is degenerate with the value of $\delta$. 
 
\begin{figure}
    \centering
    \includegraphics[width=0.48\textwidth]{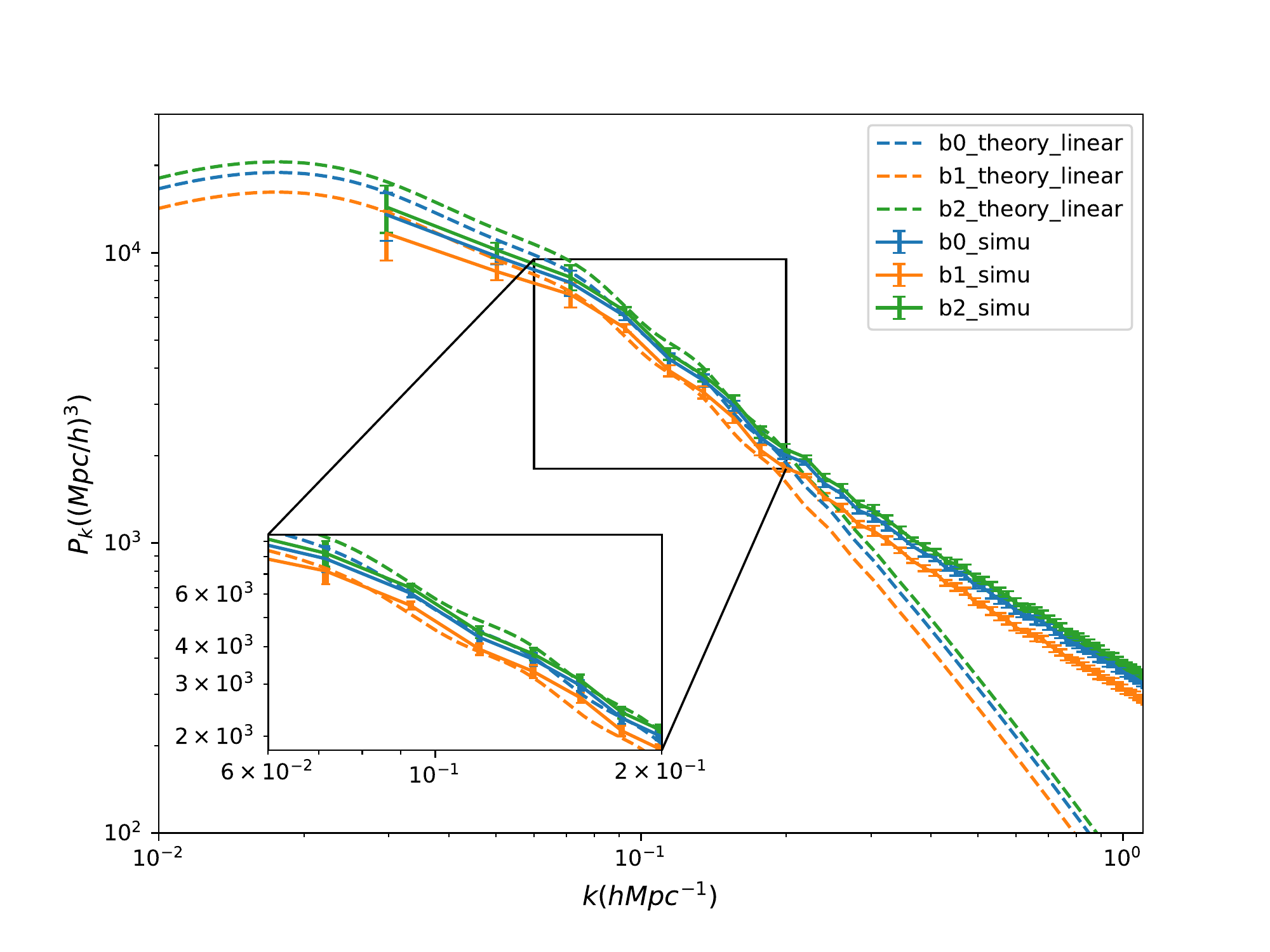}
        \caption{    {\it{ \label{Fig:pk_compare}
        Comparison between the 
simulated nonlinear power spectrum and the theoretical linear power spectrum. 
The consistency between the linear and nonlinear region is zoomed in the 
sub-graph.}}}
\end{figure}

 Finally, in Fig. \ref{Fig:pk_compare} we present   the comparison between the 
simulated nonlinear power spectrum and the theoretical linear power spectrum, 
under the 
$H(z)$ and $G_{eff}(z)$ evolution depicted in figures
 \ref{Fig:Hz} and \ref{Fig:Geff} of Section \ref{sec:model}. This is in 
agreement with  our prediction that the growth of the density field at 
large scales is consistent between theory and the simulation results.
The error bar in the figure is the root-mean-square of power spectrums, 
derived from several smaller boxes located at different places of the 
corresponding simulation boxes.

\subsection{Redshift-space distortion effect}

As it is known, in observations the distribution of galaxies is different from 
the real space, due to   the peculiar velocity of each galaxy, which 
is the so-called redshift-space distortion (RSD) effect. Since we cannot 
calculate the percentage of the measured velocity that arises from the Hubble 
flow or from the peculiar velocity,  the measured distance becomes inaccurate 
at an amount of $\Delta D$, since
\begin{equation}
    D=\frac{v}{H_0}=\frac{v_{Hubble}+V_{pec}}{H_0}=D_{real}+\Delta D.
\end{equation}
Hence, we have to test the difference of the power spectrum between real 
space and redshift space both in 3D and 2D. We use the Pylians python library 
\cite{2018ApJ...866..135V} 
\footnote{https://github.com/franciscovillaescusa/Pylians} to calculate the 
dark-matter over-density  $\delta $ with and without RSD, and thus  we 
can derive the power spectrum too.
 \begin{figure}[ht]
    \centering
    \includegraphics[width=0.5\textwidth]{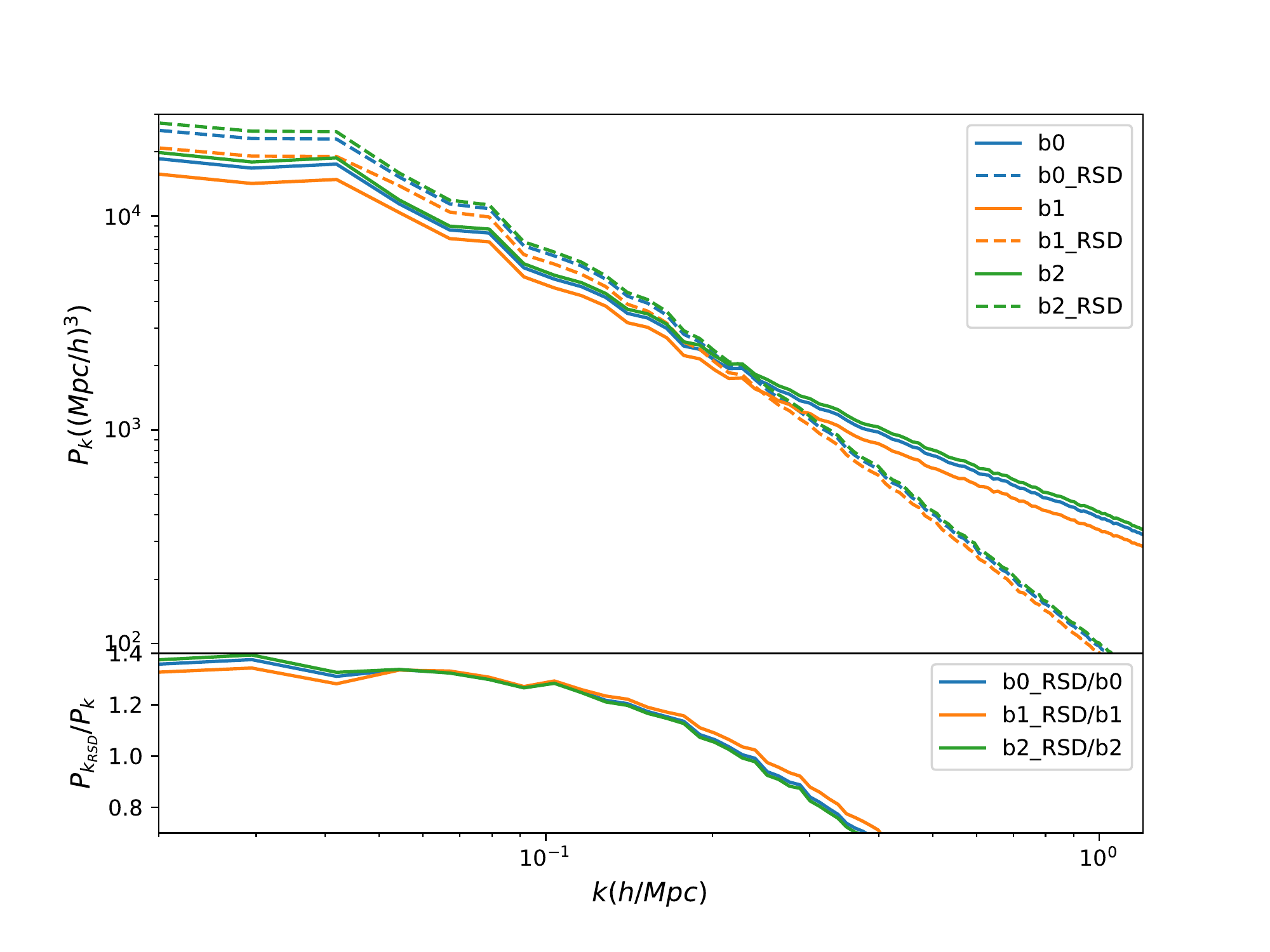}
        \caption{\label{Fig:pk_RSD}
        {\it{In the main graph we present the 
        3D power spectrum for  simulations with and without redshift space 
distortion effect, for b0 model, i.e. $\Lambda$CDM 
cosmology, and for b1 and b2 
$f(T)$ gravity models, while in the 
bottom panel we present the ratio  between them.}}}
\end{figure}

In Fig. \ref{Fig:pk_RSD} we depict  the 3D power spectrum. As we observe, the 
power spectrum is stronger in redshift space than in real space at large 
scales, but it is weaker at small scales. Furthermore, at small scales the 
scatter of the velocity of galaxies is quite large, which adds on the 
cosmological redshift along the line of sight and causes a wider distribution 
of the redshift, and this elongates the distribution of galaxies along the line 
of sight and causes a weaker correlation between galaxies, which is the so 
called ``Finger of God'' effect. Since at large scales the outside galaxies 
are still within the gravitational field and fall towards the center, then  
for the side near the observer the redshift will be larger. However, for the 
other side it will be smaller, and then the distribution of these galaxies will 
seem to be pressed and the correlation will be stronger between them. 

In the case of  b1 model, since the growth of the density field is weaker than 
b0, the scatter of the dark-matter particles is smaller,  and as it shown in 
the lower panel of   Fig. \ref{Fig:pk_RSD}, the ratio of power spectrum between 
simulations with and without RSD effect is weaker for b1 than b0, while it is 
stronger for b2 than b0. According to Fig. \ref{Fig:pk_compare} and 
Fig. \ref{Fig:pk_RSD}, we could tell that with RSD effect considered, the 
deviations between b0, b1 and b2 models are still not  so different from 
those in the real-space power spectrum.

\begin{figure}[ht]
    \centering
    \includegraphics[width=0.48\textwidth]{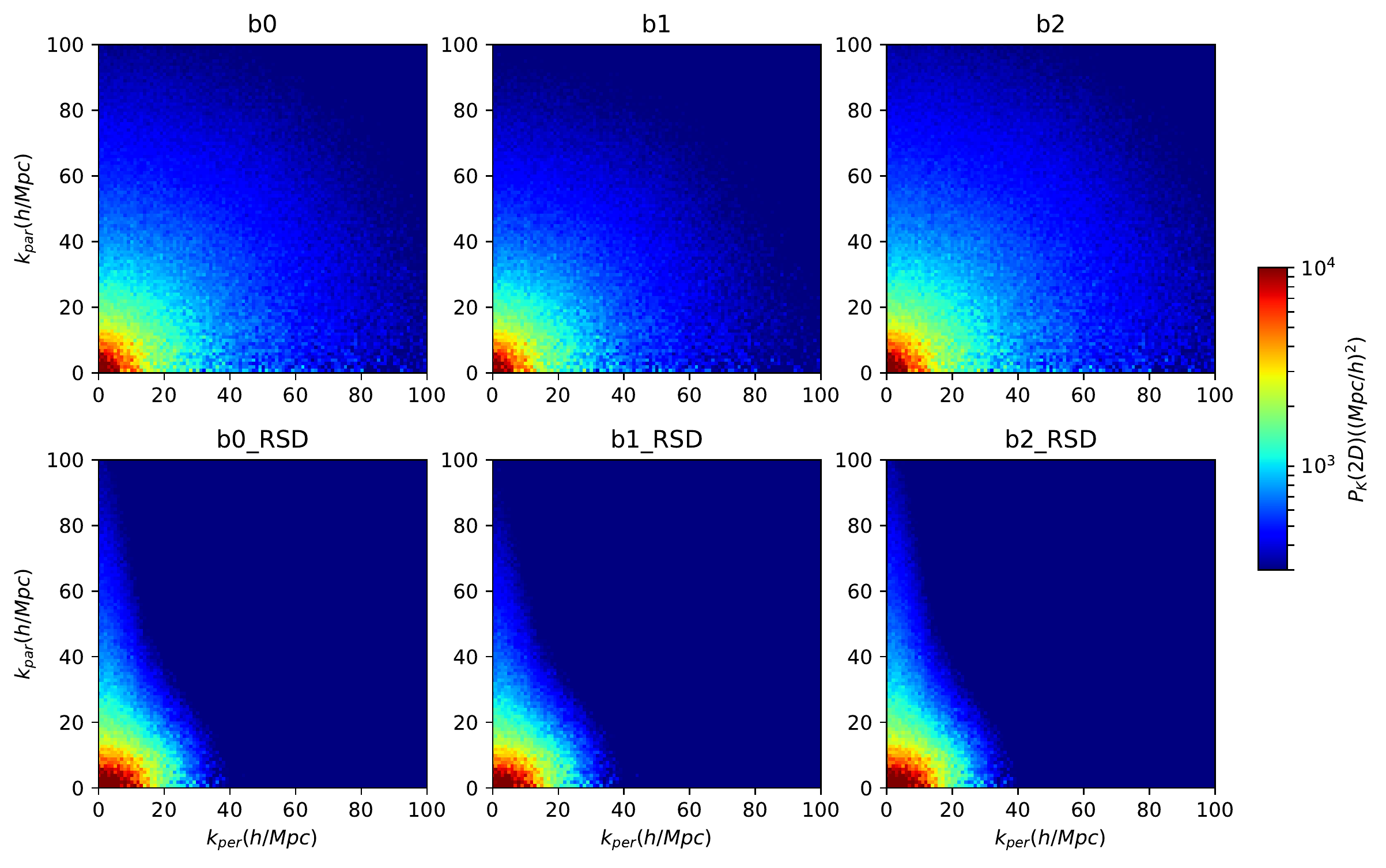}
        \caption{\label{Fig:pk_RSD_2D}
        {\it{
        2D power spectrum for  
simulations with and without redshift space distortion effect, for b0 model, 
i.e. $\Lambda$CDM 
cosmology, and for b1 and b2 
$f(T)$ gravity models.}}}
\end{figure}

  Finally, in Fig. \ref{Fig:pk_RSD_2D} we also plot the power spectrum in 2D 
to directly show the distribution in Fourier space. As we observe, the 
``Finger of God'' effect is clearly shown in the lower panel of the figure, and 
the differences between them are small.

\subsection{Counts-in-cells}

\begin{figure}
    \centering
    \includegraphics[width=0.48\textwidth]{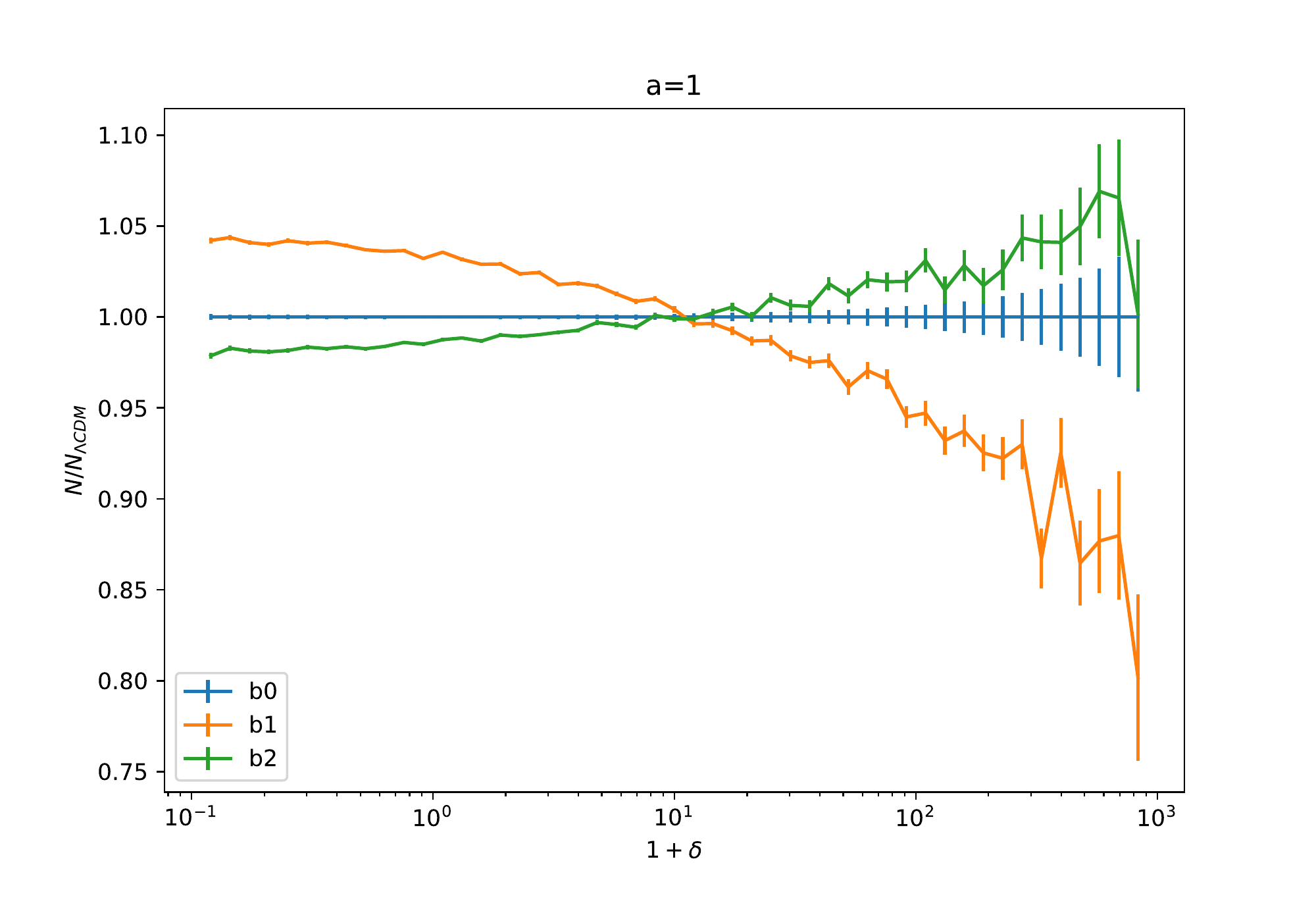}
    \caption{\label{Fig:count}
    {\it{    The ratio of over-density of cells  for b0 model, i.e. 
$\Lambda$CDM 
cosmology, and for b1 and b2 
$f(T)$ gravity models, at   present time 
$a=1$. The error bars show the Poisson noise.}}}
\end{figure}

Apart from  the correlation of  densities of each model, we can additionally 
compare the simulation particle number counts-in-cells among different 
models. The number of cells counted by over-density is shown in 
Fig. \ref{Fig:count}. 
Firstly, we calculate the over-density in $512^3$ cells for each model, and 
then we take the ratio between b1(b2) and b0 models. The error bars represent 
the Poisson error, which is estimated as $1/\sqrt{N}$. 
As expected, with larger structure growth in b2 model  comparing to 
b0 (i.e. $\Lambda$CDM) model, there are less cells with $1+\delta<10$ and more 
cells with $1+\delta>10$. On the contrary, since the structure growth in b1 
model is suppressed, there are more cells with $1+\delta<10$ and less cells 
with $1+\delta>10$. This difference indicates that with mass map reconstructed 
from weak-lensing convergence map, it is possible to distinguish $f(T)$ 
gravity from $\Lambda$CDM scenario. We also expect that the difference can be 
identified in the halo mass function. And this is what we will describe in the 
following.

\subsection{Halo mass function}

The dark-matter halo is the basic feature of large scale structure. Galaxies 
formed inside dark matter halos and sub halos. The statistics of dark matter 
halos has  been used for a long time to investigate cosmological scenarios. The 
halo mass function  is the measurement of the abundance of halos in 
different mass ranges.

\begin{figure}
    \centering
    \includegraphics[width=0.48\textwidth]{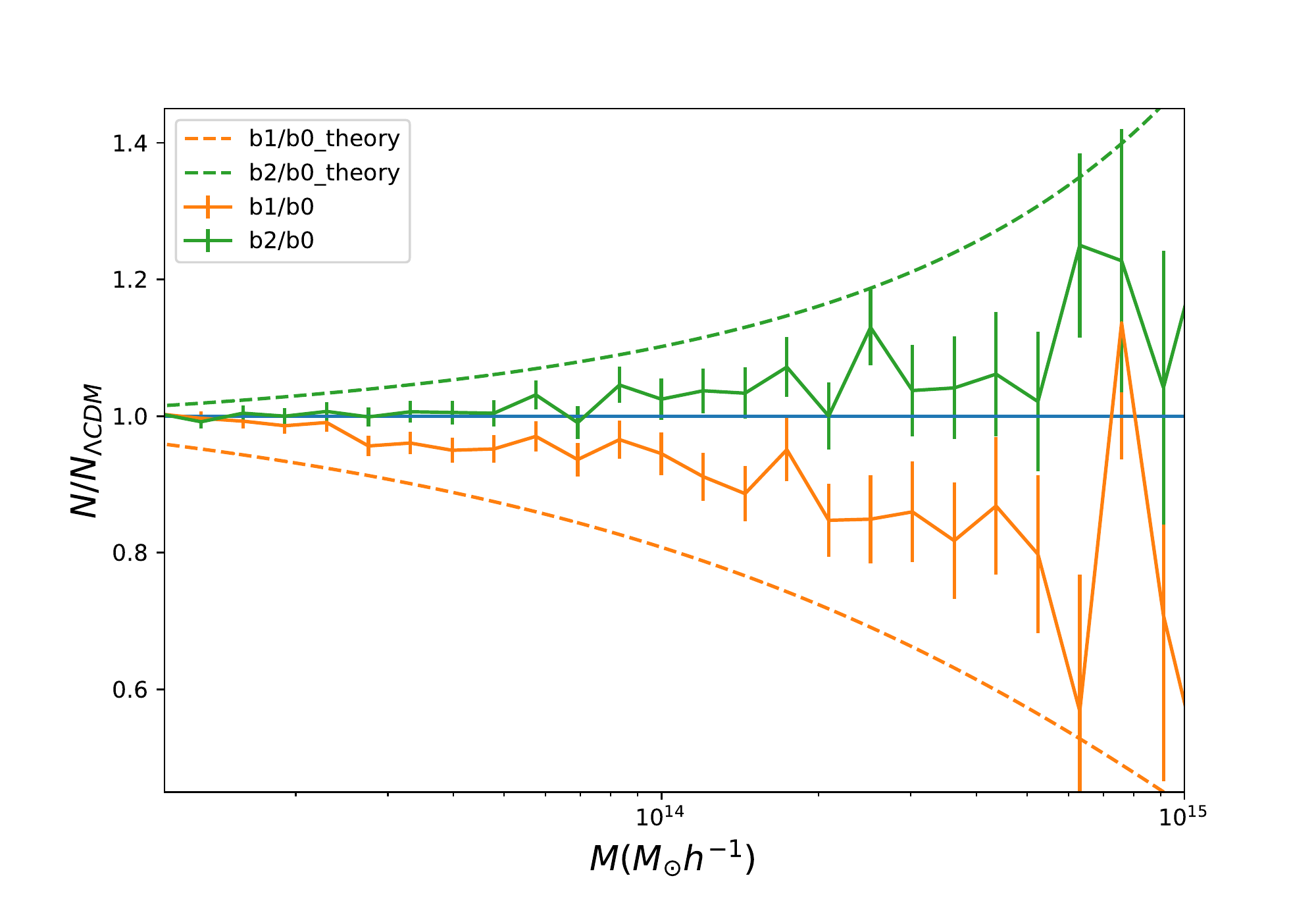}
    \caption{\label{Fig:7}
    {\it{
    The ratio of halo mass number counts (the same as 
ratio of halo mass function) between b1 and b2 models $f(T)$ gravity and b0 
model, i.e. 
$\Lambda$CDM 
cosmology. The error bars represent the Poisson noise.}}}
\end{figure}

In Fig.  \ref{Fig:7} we present 
 the ratio of number of halos in different mass ranges, resulted from our 
simulations.
As we observe, there are clearly more massive halos in b2 model than in
b0 one (i.e. $\Lambda$CDM scenario), and less massive halos in b1 model than b0 
one, which is consistent with the other observables we have discussed above. 
The number difference is very clear for halos with mass larger than 
$10^{14}M_{\odot}/h$, since clusters lie in such massive dark matter halos. 
The difference of halo mass function can lead to different predictions for 
cluster number counting, which can be compared to observations. The 
distinguishable difference of halo mass function indicates that cluster number 
counting can also be used to constrain $f(T)$ gravity. This is one of the main 
results of the present work.

After calculating the linear growth rate, it is straightforward to calculate 
the halo mass function using EPS (Extended Press-Schechter) theory:
\begin{equation}   
n(M)=\sqrt{\frac{2}{\pi}}\frac{\bar\rho}{M^2}\frac{\delta_c}{\sigma(M)}
\left|\frac {
d\ln\sigma(M)}{d\ln M}\right|\exp\left[-\frac{\delta^2_c}{2\sigma^2(M)}\right] ,
\end{equation}
where $\delta_c$ is the critical matter contrast of around 1.69. The result of 
the  simulation   is shown with dashed lines in Fig. \ref{Fig:7}. 
As we can see,   the  halo mass function   difference measured 
from simulations is not as much as that using EPS theory. It is possibly caused 
by the screening effect, in this figure, and the difference is given under 
$k_{screen}$=0.05, implying that the effect of $G_{eff}$ has been screened for 
these halos. On the other hand, the EPS theory using linear power-spectrum 
calculation, it does take $G_{eff}$ into account but not the screening, and 
hence the difference shown in EPS theory can be seen as an upper limit. This 
feature also explains why at high-mass end      we can see larger 
differences, while at low-mass end there is almost no difference.

\begin{figure*}
    \centering
   \hspace{-1cm} \includegraphics[width=1.05\textwidth]{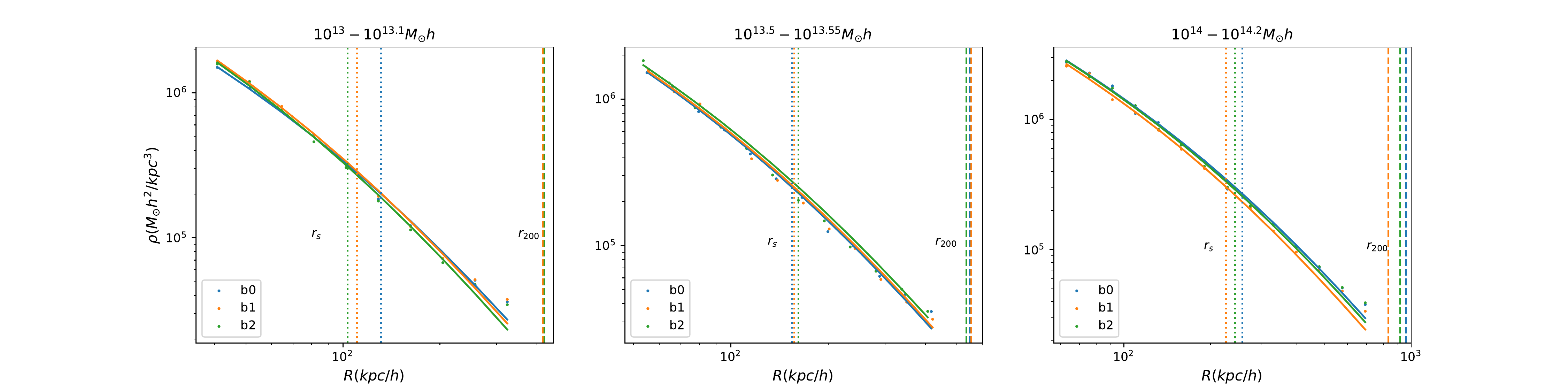}    
    \caption{\label{Fig:haloprofile}
    {\it{
    Halo density profiles for b0 model, i.e. $\Lambda$CDM 
cosmology, 
and for b1 and b2 
$f(T)$ gravity models, for three mass bins, namely
$10^{13}-10^{13.1}M_{\odot}h$, $10^{13.5}-10^{13.55}M_{\odot}h$, and 
$10^{14}-10^{14.25}M_{\odot}h$ respectively. The data points are measured from 
simulations and the curve associated with each set of points is the 
corresponding Navarro-Frenk-White (NFW) profile fitting. }}}
\end{figure*}

\subsection{Halo density profile}

In this subsection, we analyse the density profile of dark-matter halos at 
present time z=0. In particular, we measure the density profile for three 
different mass bins, namely  $10^{13}-10^{13.1}M_{\odot}h$, 
$10^{13.5}-10^{13.55}M_{\odot}h$, and $10^{14}-10^{14.25}M_{\odot}h$. We pick 
up 
the sub-halos for each mass bin, and then we count the particle numbers in each 
radial bin ranges from 0.01$r_{200}$ to $r_{200}$ kpc/h 
logarithmically ($r_{200}$ is the viral radius within which the mean density is 
200 times the cosmic matter density $\rho_m(z)$ at each redshift $z$).

In our analysis we adopt the Navarro-Frenk-White (NFW) functional form 
 \citep{1997ApJ...490..493N}, which has been an excellent 
description for the $\Lambda$CDM halo profile over a wide mass range,
given by
\begin{equation}
    \rho(r)=\frac{\rho_s}{\frac{r}{r_s}(1+\frac{r}{r_s})^2},
\end{equation}
where $\rho(r)$ is the spherically averaged density at radius $r$. There are 
two parameters in this model: $r_s$ is the scale radius and $\rho_s$ is the 
characteristic density. In Fig. \ref{Fig:haloprofile} we present the 
obtained results. As we can see,  the halo profile of b1 and b2 models of 
$f(T)$ 
gravity, as well as of  b0  $\Lambda$CDM 
cosmology, is similar in all   three mass bins. Hene, we deduce that  
the impact of $f(T)$ gravity is negligible within the halo scale.

\subsection{Low density position (LDP) lensing}
\begin{figure*}
    \centering
    \includegraphics[width=1.0\textwidth]{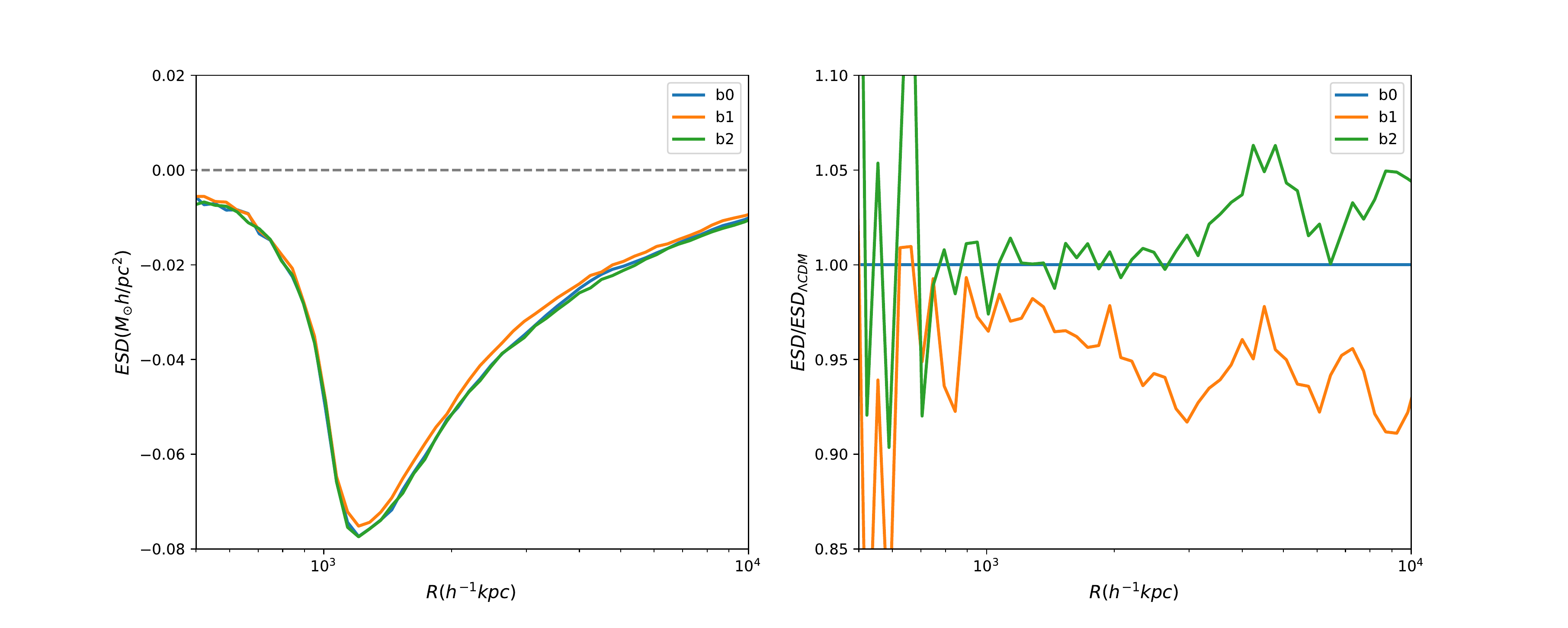}
    \caption{\label{Fig:8}
    {\it{ Left graph: Excess surface density (ESD) of LDPs for    b0 model, 
i.e. $\Lambda$CDM 
cosmology, and for b1 and b2 
$f(T)$ gravity models,  at present time $a=1$. Right  graph: 
The ratio of ESD between b1, b2 and b0 respectively.}}}
\end{figure*}

Gravitational lensing is a well-studied field in cosmology that can lead to the 
reconstruction of the matter along the line of sight. Since voids are less 
effected by the nonlinear structure formation and other physical processes, 
differences between dark energy and modified gravity models for cosmic 
acceleration might be more sensitive in cosmic voids 
\cite{2015MNRAS.450.3319L}. On the other hand, due to the screening mechanism, 
the difference between modified gravity and $\Lambda$CDM scenario is expected 
to be 
suppressed in small-scale high-density regions, such as halos. These features 
make voids a   suitable place to test modified gravity. 

The excess surface density (ESD) is related to the average background 
tangential 
shear signals through 
\begin{equation}
\Delta \Sigma(R) = \Sigma(<R)-\Sigma(R) = \gamma_t\Sigma_{crit}(z_l,z_s)   , 
\end{equation}
where 
\begin{equation}
\Sigma_{crit}=\frac{c^2}{4\pi G}\frac{D_s}{D_{ls}D_l},    
\end{equation}
and with  $D_s$, $D_l$ and $D_{ls}$   the angular diameter distance of the 
source, lens and the distance between them, respectively. In the simulations 
we can simply measure the ESD around voids.

We use the LDP defined in \cite{2019ApJ...874....7D}  to 
find the voids. In the simulation box at present time $a=1$, we exclude all 
regions of $500kpc/h$ around the most massive 7000 halos. For the left regions, 
we 
choose the grid points spaced 200kpc/h each as the LDP positions. By stacking 
the simulation particles in cylinders with $10Mpc/h$ thickness around these 
LDPs, we measure the ESD signal. The result is shown in Fig. \ref{Fig:8}. 
Again, 
this is consistent with the   measurements discussed in the previous 
subsections, and  the voids are deeper in b2 model while they are more shallow 
in b1 model. The difference is mildly distinguishable at $R>3Mpc/h$.

However,  we have to mention here  that the difference of matter distribution 
is not the only contribution to the weak lensing signal. Since the photons 
are not affected by the gravitational field in the same way as massive 
particles 
in $f(T)$ gravity, there will be some additional modification of the 
predictions. In the analysis of \citet{chen2020ft} it is shown that     
such additional modification plays an important role in galaxy-galaxy lensing. 
Thus, it would also be essential for us to include it in the LDP lensing. We 
would need to perform a more detailed analysis in order to acquire it comparing
to observations. Nevertheless, our simulation results here provided a 
self-consistent prediction of matter distribution, which forms the missing 
brick.

\section{Summary}
\label{sec:summary}

In this work, for the first time,  we performed     N-body simulations for 
$f(T)$   gravity  using ME-Gadget code,   in order to investigate in detail the 
  structure formation process. We focused on the most widely used power-law 
model, and concerning the model-parameter $b$ we considered the extreme values 
arising from cosmological observations, as well as the value $b=0$ which 
reproduces $\Lambda$CDM scenario.

Our analysis shows that  there are clear 
observational differences between $\Lambda$CDM paradigm and $f(T)$ gravity in 
the 
phenomenology of structure formation. In particular, the effect of $f(T)$ 
gravity is twofold: Firstly, due to the modifications in the Friedmann 
equations, the Hubble function (shown in Fig. \ref{Fig:Hz}) deviates from the 
one of $\Lambda$CDM cosmology, mainly around $z=1$, even for model-parameter 
values that are consistent within 1$\sigma$ with all other cosmological 
datasets (SNIa, BAO, CMB, CC etc). Secondly, since in $f(T)$ gravity the 
effective 
Newton's constant (shown in Fig. \ref{Fig:Geff})  deviates from the standard 
one, this will alter the Jeans equation and thus the growth of structure 
comparing to $\Lambda$CDM scenario. These two effects determine the deviations 
of the N-body simulations of $f(T)$ gravity from $\Lambda$CDM paradigm.

As we saw, changing the model-parameter $b$, we obtain a faster expansion 
with smaller $G_{eff}$ for $b=b1=0.20579$, and a slower expansion with larger 
$G_{eff}$ for $b=b2=-0.2371$. When $b=b0=0$  the model reproduces GR 
and 
$\Lambda$CDM cosmology. In comparison to $\Lambda$CDM model, the structure 
growth 
in 
b1 model is  slower, while the structure growth in b2 model is 
  faster. In particular, we have extracted the matter density distribution, 
matter power spectrum, counts-in-cells, halo mass function and excess surface 
density (ESD) around LDPs at present time $a=1$. Clearly, the 
simulation results are consistent with our expectations. The comparison results 
are summarized 
as follows:

\begin{itemize}
\item The matter power spectrum in b1(b2) model is smaller(larger) than 
$\Lambda$CDM  scenario, specifically   the differences
are around  $5\%\sim7.5\%$ for b1 and $10\%\sim15\%$ for b2.
In particular, we found that the contribution 
from the different expansion history is responsible  for about $2/3$ of the 
matter power spectrum difference, while the effective gravitational constant 
is responsible for the remaining $1/3$. Up to our knowledge, this is the first 
time where we acquire quantitative information on the strength of the two 
effects of the $f(T)$ modification in the structure formation.

\item There are more (less) cells with $1+\delta<10$ in b1(b2)  model, and 
less(more) cells with $1+\delta>10$ in b1(b2) model. The difference is way 
larger than the Poisson error, which may be distinguishable with weak lensing 
reconstructed mass maps.

\item There are less (more) massive halos with mass $M>10^{14}M_{\odot}/h$ in 
b1(b2) model (the difference  in halo mass function is around $\sim40\%$ at 
high mass end), and the difference is larger than the Poisson error, which may 
be distinguishable with statistical measurement of cluster number counting.

\item The ESD around LDPs is mildly different, 
nevertheless a further analysis of light-bending contribution is necessary 
in order to obtain  it compared to weak lensing observations.
\end{itemize}

In conclusion,  large scale structure can indeed lead us to distinguish $f(T)$ 
gravity from General Relativity and $\Lambda$CDM cosmology, even for ranges of 
the mode parameter that are consistent within 1$\sigma$ with cosmological 
observations. The detailed simulations of this work have set the framework for 
a quantitative discrimination and of testing General Relativity, and have 
high-lighted some possible smoking guns  of torsional modified gravity.  

One could try to use the whole analysis in order to extract 
constraints on the parameters of various  $f(T)$ models, which are expected to 
be significantly stronger than those arising from the usual cosmological 
datasets. Additionally, it would be both necessary and interesting to orient 
the investigation towards the possible alleviation of the $H_0$ and $\sigma_8$ 
tensions. These detailed studies lie beyond the scope of the present work and 
are left for   future projects.

\begin{acknowledgments}
This work was inspired  by the discussions during the second HOUYI Workshop for 
Non-standard Cosmological Models in Kunming,
China, 2019. We are grateful to Sheng-Feng Yan for helpful communications.  The 
research grants from the China Manned Space Project with no. CMS-CSST-2021-A03. 
Yi-Fu Cai is supported by the NSFC (11961131007, 11653002).
The computation resources of this work are provided by the Gravity 
Supercomputer 
at the Department of Astronomy,  Shanghai Jiao Tong University.
\end{acknowledgments}

\bibliographystyle{apsrev4-1}
\bibliography{f(T)}

\end{document}